\documentclass[prd,superscriptaddress,amsfonts,amssymb,amsmath,showpacs]{revtex4-2}
\usepackage{bm}
\usepackage{amsfonts}
\usepackage{latexsym}
\usepackage[latin1]{inputenc}
\usepackage{graphicx}
\usepackage{amsmath}
\usepackage{palatino}
\usepackage{mathpazo}
\usepackage{textcomp}
\usepackage{float}
\usepackage{booktabs}
\usepackage{dcolumn}
\usepackage{booktabs}
\usepackage{multirow}
\usepackage{hyperref}
\hypersetup{colorlinks,citecolor=red}
\usepackage{amsmath}
\usepackage{xcolor}
\usepackage{orcidlink}
\usepackage[caption=false]{subfig}
\usepackage{commath}

\def\jnl@style{\it}
\def\aaref@jnl#1{{\jnl@style#1}}

\def\aaref@jnl#1{{\jnl@style#1}}

\def\aj{\aaref@jnl{AJ}}                   
\def\apj{\aaref@jnl{ApJ}}                 
\def\apjl{\aaref@jnl{ApJ}}                
\def\apjs{\aaref@jnl{ApJS}}               
\def\apss{\aaref@jnl{Ap\&SS}}             
\def\aap{\aaref@jnl{A\&A}}                
\def\aapr{\aaref@jnl{A\&A~Rev.}}          
\def\aaps{\aaref@jnl{A\&AS}}              
\def\mnras{\aaref@jnl{Mon.~Not.~Roy.~Astron.~Soc.}}             
\def\prd{\aaref@jnl{Phys.~Rev.~D}}        
\def\prc{\aaref@jnl{Phys.~Rev.~C}}  
\def\prl{\aaref@jnl{Phys.~Rev.~Lett.}}    
\def\qjras{\aaref@jnl{QJRAS}}             
\def\skytel{\aaref@jnl{S\&T}}             
\def\ssr{\aaref@jnl{Space~Sci.~Rev.}}     
\def\zap{\aaref@jnl{ZAp}}                 
\def\nat{\aaref@jnl{Nature}}              
\def\aplett{\aaref@jnl{Astrophys.~Lett.}} 
\def\apspr{\aaref@jnl{Astrophys.~Space~Phys.~Res.}} 
\def\physrep{\aaref@jnl{Phys.~Rep.}}      
\def\physscr{\aaref@jnl{Phys.~Scr}}       
\def\commat{\aaref@jnl{Comm.~Math.~Phys.}}              
\def\science{\aaref@jnl{Science}}               
\def\cqg{\aaref@jnl{Classical Quant.~Grav.}}            
\def\jpcs{\aaref@jnl{JPCS}}                                     
\def\ijmpd{\aaref@jnl{Int.~J.~Mod.~Phys.~D}}                    
\def\grg{\aaref@jnl{Gen.~Relat.~Gravit.}}               
\def\rpp{\aaref@jnl{Rep.~Prog.~Phys.}}          
\def\npa{\aaref@jnl{Nucl.~Phys.~A}}        
\def\lrr{\aaref@jnl{Living Rev.~Rel.}}                   
\def\jcap{\aaref@jnl{J.~Cosmology Astropart.~Phys.}}    
\def\rmp{\aaref@jnl{Rev.~Mod.~Phys.}}   
\def\epjc{\aaref@jnl{Eur.~Phys.~J.~C}}

\newcommand{\udt}[3]{#1^{#2}_{\phantom{#2}#3}}

\allowdisplaybreaks[1]

\begin{document}

\color{black}       
\title{\bf Dynamical systems analysis in \texorpdfstring{$f(T,\phi)$}{} gravity}

\author{L.K. Duchaniya\orcidlink{0000-0001-6457-2225}}
\email{duchaniya98@gmail.com}
\affiliation{Department of Mathematics, Birla Institute of Technology and Science-Pilani, Hyderabad Campus, Hyderabad-500078, India.} 

\author{S.A. Kadam\orcidlink{0000-0002-2799-7870}}
\email{k.siddheshwar47@gmail.com}
\affiliation{Department of Mathematics, Birla Institute of Technology and Science-Pilani, Hyderabad Campus, Hyderabad-500078, India.}

\author{Jackson Levi Said\orcidlink{0000-0002-7835-4365}}
\email{jackson.said@um.edu.mt}
\affiliation{Institute of Space Sciences and Astronomy, University of Malta, Malta, MSD 2080}
\affiliation{Department of Physics, University of Malta, Malta}

\author{B. Mishra\orcidlink{0000-0001-5527-3565} }
\email{bivu@hyderabad.bits-pilani.ac.in }
\affiliation{Department of Mathematics, Birla Institute of Technology and Science-Pilani, Hyderabad Campus, Hyderabad-500078, India.}

\begin{abstract}
Teleparallel based cosmological models provide a description of gravity in which torsion is the mediator of gravitation. Several extensions have been made within the so-called Teleparallel equivalent of general relativity which is equivalent to general relativity at the level of the equations of motion where attempts are made to study the extensions of this form of gravity and to describe more general functions of the torsion scalar $T$. One of these extensions is $f(T,\phi)$ gravity; $T$ and $\phi$ respectively denote the torsion scalar and scalar field. In this work, the dynamical system analysis has been performed for this class of theories to obtain the cosmological behaviour of a number of models. Two models are presented here with some functional form of the torsion scalar and the critical points are obtained. For each critical point, the stability behaviour and the corresponding cosmology are shown. Through the graphical representation, the equation of state parameter and the density parameters for matter-dominated, radiation-dominated and dark energy phase are also presented for both the models.  
\end{abstract}

\maketitle

\section{Introduction} \label{SEC I}

Following supernovae cosmological observation over the last few decades \cite{Riess:1998cb,Perlmutter:1998np} several proposals have been introduced to modify the General Relativity (GR) and to consider different formulations of gravity. In a similar fashion, numerous proposals have been made to modify the Teleparallel equivalent of general relativity (TEGR) which is equivalent to GR at the level of the equations of motion. Teleparallel gravity, which is based on torsion, introduces an analogous description of gravity \cite{1985FoPh...15..365I,c_pellegrini_1963_19165,Hayashi:1979qx,Maluf:1994ji,deAndrade:2000nv,Arcos:2004tzt,Pereira:2013qza,Pereira:2019woq} in which torsion mediates gravitation. The Lagrangian density of TEGR is proportional to the torsion scalar $T$, as in this approach, the usual metric tensor and Levi-Civita connection is replaced respectively by the tetrad field and spin connection pair, and the teleparallel connection \cite{Pereira:2013qza,Arcos:2004tzt} respectively. So, the curvature and tensor based gravitational theories are equivalent at the level of the dynamical equations \cite{Arcos:2004tzt,Aldrovandi:2013wha}. Substituting the torsion scalar $T$ with the arbitrary function $f(T)$, a generalization to TEGR can be obtained \cite{Ferraro:2008ey,Bengochea:2008gz,Linder:2010py,Basilakos:2013rua,Finch:2018gkh,Bahamonde:2017ize,Basilakos:2018arq} to produce new models of cosmology. The dynamical objects in this framework are the four linearly independent tetrad fields that serves as the orthogonal basis for the tangent space at each point of the space time. The first derivative of the tetrad product was also used to construct the torsion tensor. The tetrad fields serve as dynamical variable of the teleparallel gravity, and the field equations are created by varying the action with respect to the tetrad fields. The spin connection is responsible for preserving the local Lorentz invariance of the theory and also produces  equations of motion. For more on $f(T)$ gravity, one can refer \cite{Wu:2010xk,Dent:2010nbw, Farrugia:2016qqe,Cai:2019bdh,Briffa:2020qli,LeviSaid:2021yat,Duchaniya:2022rqu}. There is strong impetus to study the generalisation with the use of scalars associated with the theoretical foundations. The non-minimal coupled scalar-torsion theory is an extension of teleparallel gravity \cite{Basilakos:2013rua,Bahamonde:2017ize} as in the case of scalar-tensor theories. This is a different class of gravitational modifications. This is because, at the level of field equations, TEGR coincides with GR, but the non-minimal coupled scalar-torsion theory does not coincide with its counterpart based on curvature.

The accelerating expansion of the Universe is sourced by some form of dark energy which takes on the form of a cosmological constant in the concordance model but is motivated by other means in modified theories of gravity \cite{Riess:1998cb,Perlmutter:1998np,WMAP:2003elm,Planck:2015bue,Planck:2018vyg}. Here, dark energy is embodied by the constant $\Lambda$ which together with GR and cold dark matter (CDM) constitute the $\Lambda$CDM model; however this model suffers with the fine tuning issue among other problems \cite{Peebles:2002gy}. One can address this issue in GR by altering the matter content of the Universe with the inclusion of additional fields such as phantom scalar, canonical scalar and vector fields \cite{Copeland:2006wr,Bassett:2005xm,Cai:2009zp}. One can also study the cosmological behaviour by creating a dynamical dark energy sector with the inclusion of scalar field such as quintessence \cite{Wetterich:1987fm,Tsujikawa:2013fta}, k-essence \cite{Chiba:1999ka,Armendariz-Picon:2000ulo}, Galileons \cite{Nicolis:2008in,Baker:2017hug,Sakstein:2017xjx} and so on. Another way to address this accelerated expansion issue is by extending or altering the geometrical part of Einstein-Hilbert action, that leads to different extended/modified theories of gravity.

The dark energy scenario was studied in the scalar-torsion theory with non-minimal coupling between torsion scalar and dynamical scalar field in (Geng et al. \cite{Geng:2011aj,Geng:2011ka}). Similar study was made with an arbitrary non-minimal coupling function and tachyon term for scalar field \cite{Otalora:2013dsa,Otalora:2013tba}. One of the extensions of $f(T)$ gravity is the generalised scalar-torsion $f(T,\phi)$ gravity, where $\phi$ is the canonical scalar and in the gravitational action the scalar field is non-minimally connected with torsion scalar \cite{Xu:2012jf}. Further in the covariant teleparallel framework, a new class of theories have been given where the action depends on the scalar field and arbitrary function  of torsion scalar \cite{Hohmann:2018rwf}. M. G. Espinoza and G. Otalora \cite{Gonzalez_Espinoza_2020A} have studied the generation of primordial fluctuations in generalized teleparallel scalar-torsion gravity theories whose Lagrangian density is an arbitrary function $f(T, \phi)$ of the torsion scalar $T$ and a scalar field $\phi$, plus the kinetic term and develop primordial density
perturbations started from the Arnowitt-Deser-Misner
(ADM) formalism of the tetrad field. The cosmological dynamics of dark energy and its stability was studied in \cite{Gonzalez-Espinoza:2020jss} and the scalar perturbation was done in \cite{Gonzalez-Espinoza:2021mwr}. Several models were set in the context of the dynamical system to reveal the evolutionary behaviour of the dark energy models \cite{Kadam:2022lgq}. Motivated with this non-minimal coupling of torsion scalar and scalar field, in this paper we will study the cosmological aspects of the models through the dynamical system analysis. 

The paper is organised as follows, in Sec.~\ref{SECII}, the basic equations of teleparallel gravity and the field equations of of $f(T,\phi)$ gravity in an isotropic an homogeneous space time are given. In Sec.~\ref{SECIII}, the dynamical system analysis for two models are performed and the evolutionary behaviour of the models are studied. In Sec.~\ref{SECIV}, the discussions and conclusions of the models are presented.

\section{Field Equations of the scalar-torsion \texorpdfstring{$f(T,\phi)$}{} gravity}\label{SECII}

The TEGR action is composed simply of the linear torsion scalar, which can be immediately generalized to $f(T)$ gravity. By elevating the torsion scalar to an arbitrary function thereof, the addition of a scalar field $\phi$ can be introduced by writing the action as \cite{Gonzalez-Espinoza:2021mwr}
\begin{equation}\label{1}
    S =\int d^{4}xe[f(T,\phi)+P(\phi)X]+S_{m}+S_{r}\,,
\end{equation}
where $e = \det[e^A_{\mu}] = \sqrt{-g}$ is the determinant of the tetrad field. Matter action is denoted by the symbol $S_{m}$, whereas radiation action is characterized by $S_{r}$. Using tetrad and spin connection pair as the dynamical variable in place of metric tensor, GR can also be expressed in the framework of teleparallel gravity. The tetrad field, $e^{A}_{\mu}$, $A = 0,1,2,3$, metric tensor $g_{\mu \nu}$ and the Minkowski tangent space metric $\eta_{AB}$ can have the local relation as, $g_{\mu \nu}=\eta_{AB} e_{\mu}^{A} e_{\nu}^{B}$, where $\eta_{AB}=(-1,1,1,1)$. The tetrad satisfies the orthogonality condition, $e^{\mu}_Ae^B_{\mu}=\delta_A^B$, whereas the spin connection is denoted by $\udt{\omega}{A}{B\mu}$. The function $f(T,\phi)$ represents an arbitrary function of scalar field $\phi$ and the torsion scalar $T$ and $X= -\partial_{\mu} \phi \partial^{\mu} \phi/2$ is the kinetic term of the field. Non-minimally coupled scalar-torsion gravity models with the coupling function $f(T, \phi)$, $f(T)$ gravity and minimally coupled scalar field are all included in this general action. The torsion scalar is
\begin{equation}\label{2}
    T = S_{\theta}^{\mu \nu} T_{\mu \nu}^{\theta}\,,
\end{equation}
where $S_{\theta}^{\mu \nu}$ and $T_{\mu \nu}^{\theta}$ respectively represents the superpotential and the torsion tensor. Further, the superpotential can be expressed as,
\begin{equation} \label{3}
    S_{\theta}^{~~\mu \nu}\equiv\frac{1}{2}(K^{\mu \nu}_{~~~\theta}+\delta^{\mu}_{\theta}T^{\alpha \nu}_{~~~\alpha}-\delta^{\nu}_{\theta}T^{\alpha \mu}_{~~~\alpha})\,,
\end{equation} 
where $K^{\mu \nu}_{~~~\theta}\equiv \frac{1}{2}(T^{\nu \mu}_{~~~\theta}+T_{\theta}^{~~\mu \nu}-T^{\mu \nu}_{~~~\theta})$ be the contortion tensor. The torsion tensor is represented by
\begin{equation}\label{4}
    T_{\mu \nu}^{\theta} = e^{\theta}_{A}\partial_{\mu} e^{A}_{\nu}-e^{\theta}_{A}\partial_{\nu}e^{A}_{\mu}+e^{\theta}_{A} \omega^{A}_{B\mu}e^{B}_{\nu}-e^{\theta}_{A} \omega^{A}_{B\nu}e^{B}_{\mu}\,.
\end{equation}
There also exists special frames in which the spin connection vanishes, which is known as the Weitzenb$\ddot{o}$ck gauge.

Now, the gravitational field equations can be obtained either varying the action with respect to the tetrad $e^{A}_{\mu}$ or with the relation between the curvature and torsion scalar with the use of Levi-Civita connection and contortion tensor to obtain \cite{Hohmann:2018rwf}
\begin{equation}\label{6}
    T = -R + 2 e^{-1}\partial_{\mu}(eT^{\alpha\mu}_{~~\alpha})\,.
\end{equation}
In order to obtain the field equations of $f(T,\phi)$ gravity to study its cosmological applications, we consider the homogeneous and isotropic flat Friedmann-Lema\^{i}tre-Robertson-Walker (FLRW) space time as
\begin{equation}\label{7}
    ds^{2}=-dt^{2}+a^{2}(t)[dx^2+dy^2+dz^2]\,,
\end{equation}
where $a(t)$ is the scale factor that represents the expansion rate in the spatial directions and the tetrad, $e^{A}_{\mu} = diag(1,a,a,a)$. Varying the action in Eq.~\eqref{1} with respect to the tertad field and the scalar field $\phi$, we can obtain the equations of motion of $f(T,\phi)$ cosmology as,
\begin{eqnarray}
    f(T,\phi)-P(\phi)X-2Tf,_{T}&=&\rho_{m}+\rho_{r} \label{8}\\
    f(T,\phi)+P(\phi)X-2Tf,_{T}-4\dot{H}f,_{T}-4H\dot{f},_{T} &=& -p_{r}\label{9}\\
    -P,_{\phi}X-3P(\phi)H\dot{\phi}-P(\phi)\ddot{\phi}+f,_{\phi}&=&0\,. \label{10}
\end{eqnarray}

The Hubble parameter $H\equiv\frac{\dot{a}}{a}$ with an over dot denotes the derivative with respect to the cosmic time $t$. We represent $f\equiv f(T,\phi)$ and $f_{,T}=\frac{\partial f}{\partial T}$. The energy density for matter and radiation are denoted as $\rho_{m}$, $\rho_{r}$ respectively and the pressure at radiation era is $p_{r}$. From Eq.~\eqref{2}, one can obtain the torsion scalar, $T=6H^{2}$. In Eqs.~\eqref{8}--\eqref{10}, we consider the non-minimal coupling function $f(T,\phi)$ in the form \cite{Hohmann:2018rwf}
\begin{equation}\label{11}
    f(T,\phi)=-\frac{T}{2\kappa^{2}}-G(T)-V(\phi)\,,
\end{equation}
where $V(\phi)$ is the scalar potential and $G(T)$ is the arbitrary function of torsion scalar. We consider for matter dominated era $\omega_{m}=\frac{p_{\rm m}}{\rho_{\rm m}} = 0$, and for radiation era $\omega_{r}=\frac{p_{r}}{\rho_{\rm r}} = 1/3$, subsequently Eqs.~\eqref{8}--\eqref{10} reduce to
\begin{eqnarray}
    \frac{3}{\kappa^{2}}H^{2}=P(\phi)X+V(\phi)-2TG_{,T}+G(T)+\rho_{m}+\rho_{r}\,,\label{12}\\
    -\frac{2}{\kappa^{2}}\dot{H}=2P(\phi)X+4\dot{H}(G_{T}+2TG_{,TT})+\rho_{m}+\frac{4}{3}\rho_{r}\,,\label{13}\\
    P(\phi)\ddot{\phi}+P_{,\phi}(\phi)X+3P(\phi)H\dot{\phi}+V_{,\phi}(\phi)=0\,.\label{14}
\end{eqnarray}

The Friedmann Eqs.~\eqref{12}--\eqref{13} are then modified to give
\begin{eqnarray}
    \frac{3}{\kappa^{2}}H^{2}=\rho_{m}+\rho_{r}+\rho_{de}\,, \label{15}\\
    -\frac{2}{\kappa^{2}}\dot{H}=\rho_{m}+\frac{4}{3}\rho_{r}+\rho_{de}+p_{de}\,. \label{16}
\end{eqnarray}

Comparing Eq.~\eqref{12} with Eq.~\eqref{15}, and Eq.~\eqref{13} with Eq.~\eqref{16}, the energy density ($\rho_{de}$) and pressure ($p_{de}$) for the dark energy sector can be retrieved as,
\begin{align}
    \rho_{de} &= P(\phi)X+V(\phi)-2TG_{,T}+G(T)\,, \label{17}\\
    p_{de} &= P(\phi)X-V(\phi)+2TG_{,T}-G(T)+4\dot{H}(G_{,T}+2TG_{,TT})\,. \label{18}
\end{align}

For the sake of brevity, we take $P(\phi)$ = 1. The potential energy, $V(\phi)=V_{0}e^{-\lambda\phi}$, where $\lambda$ is a constant. Our motivation is to construct the cosmological models of the Universe in the dark energy sector along with its dynamical system analysis. In order to develop the system, the form of $G(T)$ would be needed and therefore in the subsequent section we shall consider two forms of $G(T)$ to represent two models.

\section{Dynamical System Analysis of the Models}\label{SECIII}

The motivation of this work is to study the cosmological dynamics of some models within the general class of scalar-tensor theories with nontrivial torsion scalar contributions. The dynamical system is a concept that specifies some rule for the development of the system and the possible future behaviour of the cosmological models. An equation of the form $Y^{\prime}=f(Y)$ represents a dynamical system, where $Y$ is the column vector constituted by suitable auxiliary variables and $f(Y)$ be the corresponding column vector of the autonomous equations. The prime denotes derivative with respect to $N = ln a$. This analysis helps to understand the overall dynamics of the Universe by identifying the critical points at which $f(Y)$ vanishes. We propose here two models with some popular form of $G(T)$.

\subsection{Model I}\label{sec:model_1}

For the first model, we consider $G(T)$ as \cite{2011JCAP...07..015Z}
\begin{equation}\label{19}
    G(T)= \beta T \ln \left(\frac{T}{T_{0}}\right)\,,
\end{equation}
where $\beta$ be the constant and $T_0$ be the value of $T$ at the initial epoch. This model has been shown to produce \cite{2011JCAP...07..015Z} physically advantageous critical points and may be interesting to model the evolution of the Universe. Here, the effective dark energy density and the effective dark energy pressure terms in Eqs.~\eqref{17}--\eqref{18} reduce to
\begin{eqnarray}
    \rho_{de}&=&\frac{\dot{\phi}^{2}}{2}+V(\phi)-6\beta H^{2} \ln \left(\frac{6 H^{2}}{T_{0}}\right)- 12 H^{2} \beta\,, \label{20} \\ 
    p_{de}&=&\frac{\dot{\phi}^{2}}{2}-V(\phi)+6\beta H^{2} \ln \left(\frac{6H^{2}}{T_{0}}\right)+ 12 H^{2} \beta 
    + 4 \dot{H} \left(\beta \ln \left(\frac{6H^{2}}{T_{0}} \right)+3 \beta \right)\,, \label{21}
\end{eqnarray}
and the scalar field Klein-Gordon equation \eqref{14} becomes
\begin{equation}\label{22}
    \ddot{\phi}+3H\dot{\phi}+V,_{\phi}(\phi)=0\,,
\end{equation}
which can also be written as,
\begin{equation}\label{23}
    \frac{d}{dt}\left(\frac{\dot{\phi}^{2}}{2}+V(\phi)\right)=-3H\dot{\phi}^{2}\,.
\end{equation}
Also, the fluid equation for dark energy sector can be written as 
\begin{equation} \label{24}
    \dot{\rho}_{de}+3H(\rho_{de}+p_{de})=0\,.
\end{equation} 
The density parameters for matter-dominated (${\Omega_m}$), radiation-dominated (${\Omega_r}$) and dark energy sector (${\Omega_{\Lambda}}$) can be constrained through
\begin{equation}\label{25}
    \Omega_{m}+\Omega_r+\Omega_{\Lambda}=1\,,
\end{equation}
where ${\Omega_m}=\frac{\kappa^2 \rho_m}{3H^2}$, ${\Omega_r}=\frac{\kappa^2 \rho_r}{3H^2}$, ${\Omega_{\Lambda}}=\frac{\kappa^2 \rho_{de}}{3H^2}$. From Eqs.~\eqref{20}--\eqref{21}, the equation of state parameter can be obtained as 
\begin{equation}\label{26}
    \omega_{de}\equiv\frac{p_{de}}{\rho_{de}}= \frac{\dot{\phi^{2}}-2 V(\phi)+12 H^{2} \beta \ln\left(6\frac{H^{2}}{T_{0}}\right)+24 H^{2} \beta +8 \dot{H} \left(\beta \ln (6\frac{H^{2}}{T_{0}})+3 \beta \right)}{\dot{\phi^{2}}+2 V(\phi)- 12 H^{2} \beta \ln\left (6\frac{H^{2}}{T_{0}}\right)-24 H^{2} \beta }\,.
\end{equation}

To study the dynamics of the model in scalar-torsion $f(T,\phi)$ gravity, we introduce the following dimensionless phase space variables in order to frame the autonomous dynamical system as,
\begin{align}
    x=\frac{\kappa\dot{\phi}}{\sqrt{6}H}\,, \hspace{1cm} y=\frac{\kappa\sqrt{V}}{\sqrt{3}H}\,, \hspace{1cm}
    z=-4 \beta  \kappa^{2}\,, \hspace{1cm}
    u=-2 \beta \ln \left(\frac{T}{T_{0}}\right)\kappa^{2} \,, \label{27}\\ 
    \rho=\frac{\kappa\sqrt{\rho_{r}}}{\sqrt{3}H}\,, \hspace{1cm} 
    \lambda= -\frac{V_{,\phi}(\phi)}{\kappa V(\phi)}\,, \hspace{1cm} 
    \Theta= \frac{V(\phi)\,, V_{,\phi \phi}}{V_{,\phi}(\phi)^{2}}\,. \label{28}
\end{align}
The density parameter for different phases of the evolution of the Universe in terms of dynamical system variable are as follow,
\begin{align}
    \Omega_{de}&=x^{2}+y^{2}+z+u\,, \label{29} \\ 
    \Omega_{r}&=\rho^{2}\,, \label{30} \\ 
    \Omega_{m}&=1-x^{2}-y^{2}-z-u-\rho^{2}\,, \label{31}
\end{align}
The Friedmann Eqs.~\eqref{12}--\eqref{13} and the variables in Eqs.~\eqref{27}--\eqref{28} would reproduce
\begin{eqnarray}\label{32}
    \frac{\dot{H}}{H^{2}}&=\frac{\rho ^2-3 \left(u-x^2+y^2+z-1\right)}{2 u+3 z-2}\,,
\end{eqnarray}
so that the deceleration parameter and equation of state (EoS) parameter can also be expressed in terms of dynamical variables as,
\begin{eqnarray}
    q &=&\frac{\rho ^2-u+3 x^2-3 y^2+1}{-2 u-3 z+2}\,,\\ \label{33}
    \omega_{tot}&=& \frac{2 \rho ^2+6 x^2-6 y^2+3 z}{-6 u-9 z+6}\,,\\ \label{34}
    \omega_{de}&=& -\frac{\rho ^2 (2 u+3 z)+6 x^2-6 y^2+3 z}{3 (2 u+3 z-2) \left(u+x^2+y^2+z\right)}\,. \label{35}
\end{eqnarray}
The system of autonomous equations that governs the cosmological dynamical system are
\begin{align}
    \frac{dx}{dN}&=-\frac{x\rho ^2-3 x\left(u-x^2+y^2+z-1\right)}{2 u+3 z-2}-3 x+\sqrt{\frac{3}{2}} \lambda  y^2\,, \label{36} \\ 
    \frac{dy}{dN}&=\frac{-y \rho ^2+3y \left(u-x^2+y^2+z-1\right)}{2 u+3 z-2}-\sqrt{\frac{3}{2}} \lambda y x\,, \label{37} \\
    \frac{du}{dN}&=\frac{z \rho ^2-3 z \left(u-x^2+y^2+z-1\right)}{2 u+3 z-2}\,, \label{38} \\ 
    \frac{d\rho}{dN}&=-\frac{\rho \left(\rho ^2+u+3 x^2-3 y^2+3 z-1\right)}{2 u+3 z-2}\,, \label{39} \\
    \frac{dz}{dN}&=0\,,  \label{40}\\
    \frac{d\lambda}{dN}&= -\sqrt{6}(\Theta-1)x \lambda^{2}\,. \label{41}
\end{align}

In order to derive the dynamical features of the autonomous system, the coupled equations $x^{\prime}=0$, $y^{\prime}=0$, $z^{\prime}=0$, $u^{\prime}=0$ and $\rho^{\prime}=0$ are to be solved. The special choice of the potential energy function, $V(\phi)=V_{0}e^{-\lambda\phi}$, leads to the value of $\Theta=1$. The corresponding critical points of the above system and its description are given in Table~\ref{TABLE-I}. The stability condition and the cosmology pertaining to the value of deceleration and EoS parameter are given in Table~\ref{TABLE-II}. The cosmological solution and the corresponding scale factor are also given in Table~\ref{TABLE-III}. 

\begin{table}[H]
    \caption{Critical points for the dynamical system. } 
    \centering 
    \begin{tabular}{|c|c|c|c|c|c|c|} 
    \hline\hline 
    C.P. & $x_{c}$ & $y_{c}$ & $u_{c}$ & $\rho_{c}$ & $z_{c}$ & Exists for \\ [0.5ex] 
    \hline\hline 
    $A$  & $0$ & $0$ & $\alpha$ & 0 & $\beta_{3}$ & $\begin{tabular}{@{}c@{}} $\alpha=1-\beta_{3}$,\\ $ \beta_{3} \neq 0$ \end{tabular}$ \\
    \hline
    $B$ &$0$ & $0$ & $\gamma$ & $0$&0 & $ \gamma \neq 1$ \\
    \hline
    $C$ & 0 & 0 & $\sigma$ & $\tau$ & 0 & $\tau =\sqrt{1-\sigma}$, $\sigma<1$ \\
    \hline
    $D$ & $\delta$ & 0 & $\epsilon$ & 0 & 0 & $\delta \neq 0,$ $\epsilon =1-\delta ^2 $ \\
    \hline
    $E$ & 0 & $\eta$ & $\iota$ & 0 & $\xi$ & $\begin{tabular}{@{}c@{}}$\iota =-\eta ^2-\xi +1,$\\ $2 \eta ^2-\xi \neq 0$, $\lambda=0$\end{tabular}$\\
    \hline
    $F_{+}$  & $\frac{\sqrt{\frac{3}{2}}}{\lambda }$ & $\frac{\sqrt{\frac{3}{2}}}{\lambda }$ & $\mu$ & 0 & $0$ & $\mu -1\neq 0,$ $\lambda \neq 0$\\
    \hline
    $F_{-}$  & $\frac{\sqrt{\frac{3}{2}}}{\lambda }$ & $-\frac{\sqrt{\frac{3}{2}}}{\lambda }$ & $\nu$ & 0 & $0$ & $\nu -1\neq 0,$ $\lambda \neq 0$\\
    \hline
    $G$ & 0 & $\mathbf{f}$ & $\mathbf{e}$ & 0 & 0 & $\begin{tabular}{@{}c@{}} $\mathbf{e}=1-\mathbf{f}^2,$ $\mathbf{f}\neq 0,$\\ $\lambda=0$\end{tabular}$\\
    \hline
    $\mathcal{H}$ & 0 & $\mathbf{i}$& $\mathbf{h}$ & 0 & $\mathbf{j}$& $\begin{tabular}{@{}c@{}}$\lambda \neq 0,$ $\mathbf{h}=-\mathbf{i}^2-\mathbf{j}+1,$\\ $-\mathbf{j} \lambda \neq 0$\end{tabular}$\\
    [1ex] 
    \hline 
    \end{tabular}
    \label{TABLE-I}
\end{table}

\begin{table}[H]
    \caption{Stability conditions, EoS parameter and deceleration parameter } 
    \centering 
    \begin{tabular}{|c|c|c|c|c|} 
    \hline\hline 
    C. P. & Stability Conditions & $q$ & $\omega_{tot}$ & $\omega_{de}$ \\ [0.5ex] 
    \hline\hline 
    $A$  & Stable & $-1$ & $-1$ & $-1$ \\
    \hline
    $B$  & Unstable & $\frac{1}{2}$ & 0 & 0 \\
    \hline
    $C$  &  Unstable & $1$ & $\frac{1}{3}$ & $\frac{1}{3}$ \\
    \hline
    $D$  &Unstable& $2$ & $1$ & $1$ \\
    \hline
    $E$  & Stable & $-1$ & $-1$ & $-1$ \\
    \hline
    $F_{+}$  & \begin{tabular}{@{}c@{}}Stable for \\ $\mu <1\land \left(-2 \sqrt{\frac{6}{7}} \sqrt{-\frac{1}{\mu -1}}\leq \lambda <-\sqrt{3} \sqrt{-\frac{1}{\mu -1}}\lor \sqrt{3} \sqrt{-\frac{1}{\mu -1}}<\lambda \leq 2 \sqrt{\frac{6}{7}} \sqrt{-\frac{1}{\mu -1}}\right)$ \end{tabular} & $\frac{1}{2}$ & $0$ & $0$ \\
    \hline
    $F_{-}$  & \begin{tabular}{@{}c@{}}Stable for \\ $\nu <1\land \left(-2 \sqrt{\frac{6}{7}} \sqrt{-\frac{1}{\nu -1}}\leq \lambda <-\sqrt{3} \sqrt{-\frac{1}{\nu -1}}\lor \sqrt{3} \sqrt{-\frac{1}{\nu -1}}<\lambda \leq 2 \sqrt{\frac{6}{7}} \sqrt{-\frac{1}{\nu -1}}\right)$ \end{tabular} & $\frac{1}{2}$ & $0$ & $0$ \\
    \hline
    $G$  & Stable & $-1$ & $-1$ & $-1$\\
    \hline
    $\mathcal{H}$ & \begin{tabular}{@{}c@{}}Stable for\\ $\left(\mathbf{i}<1\land \mathbf{j}>2 \mathbf{i}^2\right)\lor \left(\mathbf{i}>1\land \mathbf{j}>2 \mathbf{i}^2\right)$\end{tabular} & $\frac{\mathbf{i}^2+\mathbf{j}}{2 \mathbf{i}^2-\mathbf{j}}$ & $\frac{\mathbf{j}}{2 \mathbf{i}^2-\mathbf{j}}$ & $\frac{\mathbf{j}}{\left(\mathbf{i}^2-1\right) \left(\mathbf{j}-2 \mathbf{i}^2\right)}$ \\
    [1ex] 
    \hline 
    \end{tabular}
    \label{TABLE-II}
\end{table}

\begin{table}[H]
    \caption{Cosmological solutions of critical points } 
    \centering 
    \begin{tabular}{|c|c|c|c|} 
    \hline\hline 
    C. P. & Acceleration equation & Scale factor ( Power law solution) & Universe phase \\ [0.5ex] 
    \hline\hline 
    $A$ & $\dot{H}=0$ & $a(t)=t_{0} e^{c_{1}t}$ & de-sitter phase \\
    \hline
    $B$ & $\dot{H}=-\frac{3}{2}H^{2}$ & $a(t)= t_{0} (\frac{3}{2}t+c_{2})^\frac{2}{3}$ & matter-dominated \\
    \hline
    $C$ & $\dot{H}=-2 H^{2}$ & $a(t)= t_{0} (2 t+c_{2})^\frac{1}{2}$ & radiation-dominated \\
    \hline
    $D$ &$\dot{H}=-3 H^{2}$& $a(t)= t_{0} (3 t+c_{2})^\frac{1}{3}$ & stiff-matter \\
    \hline
    $E$ & $\dot{H}=0$ & $a(t)=t_{0} e^{c_{1}t}$ & de-sitter phase \\
    \hline
    $F_{+}$ & $\dot{H}=-\frac{3}{2}H^{2}$ & $a(t)= t_{0} (\frac{3}{2}t+c_{2})^\frac{2}{3}$ & matter-dominated \\
    \hline
    $F_{-}$ & $\dot{H}=-\frac{3}{2}H^{2}$ & $a(t)= t_{0} (\frac{3}{2}t+c_{2})^\frac{2}{3}$ & matter-dominated \\
    \hline
    $G$ &$\dot{H}=0$ & $a(t)=t_{0} e^{c_{1}t}$ & de-sitter phase \\
    \hline
    $\mathcal{H}$ & $\dot{H}=0$ & $a(t)=t_{0} e^{c_{1}t}$ & de-sitter phase \\
    [1ex] 
    \hline 
    \end{tabular}
    \label{TABLE-III}
\end{table}

\begin{figure}[H]
    \centering
    \includegraphics[width=58mm]{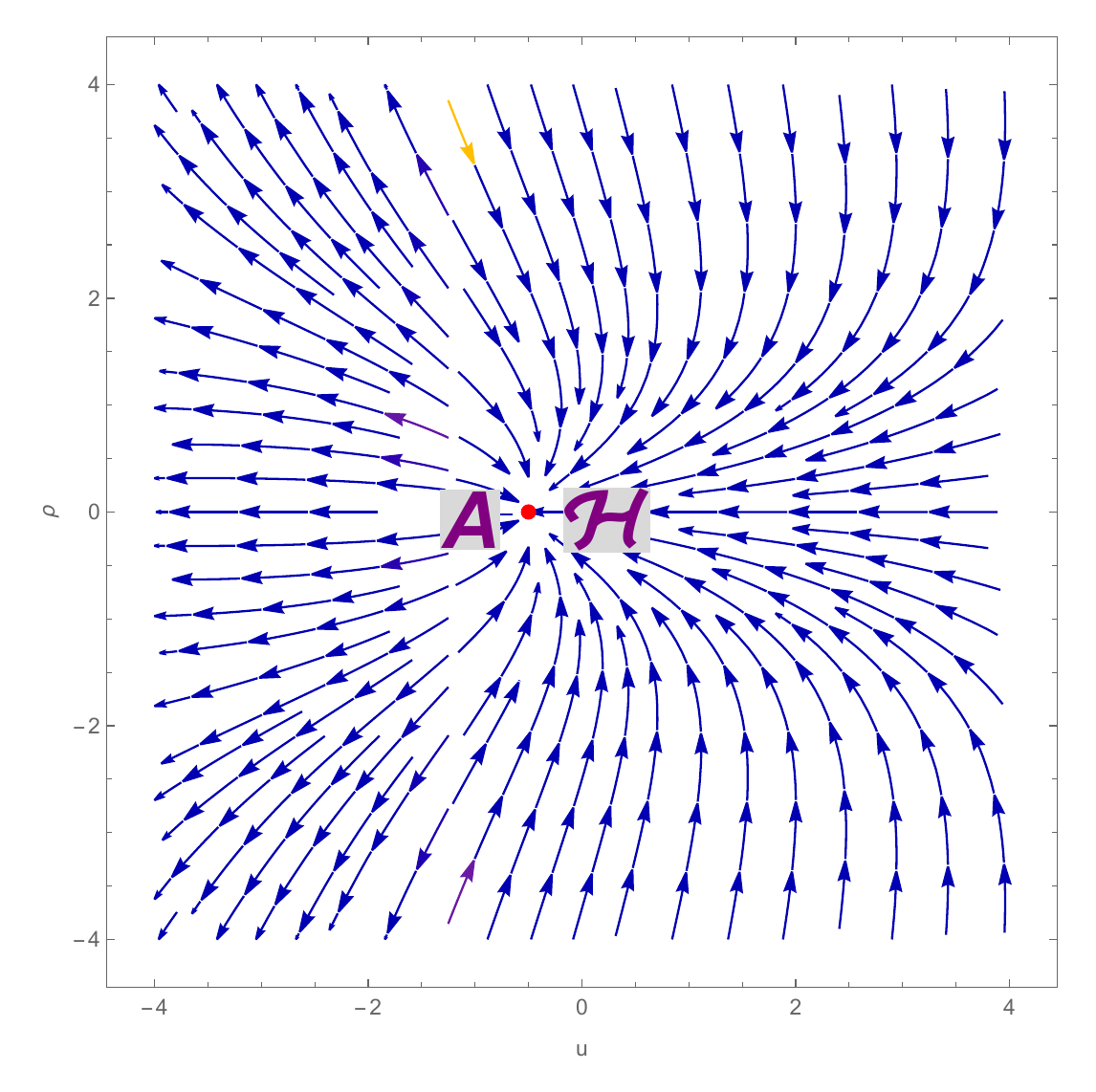}
    \includegraphics[width=58mm]{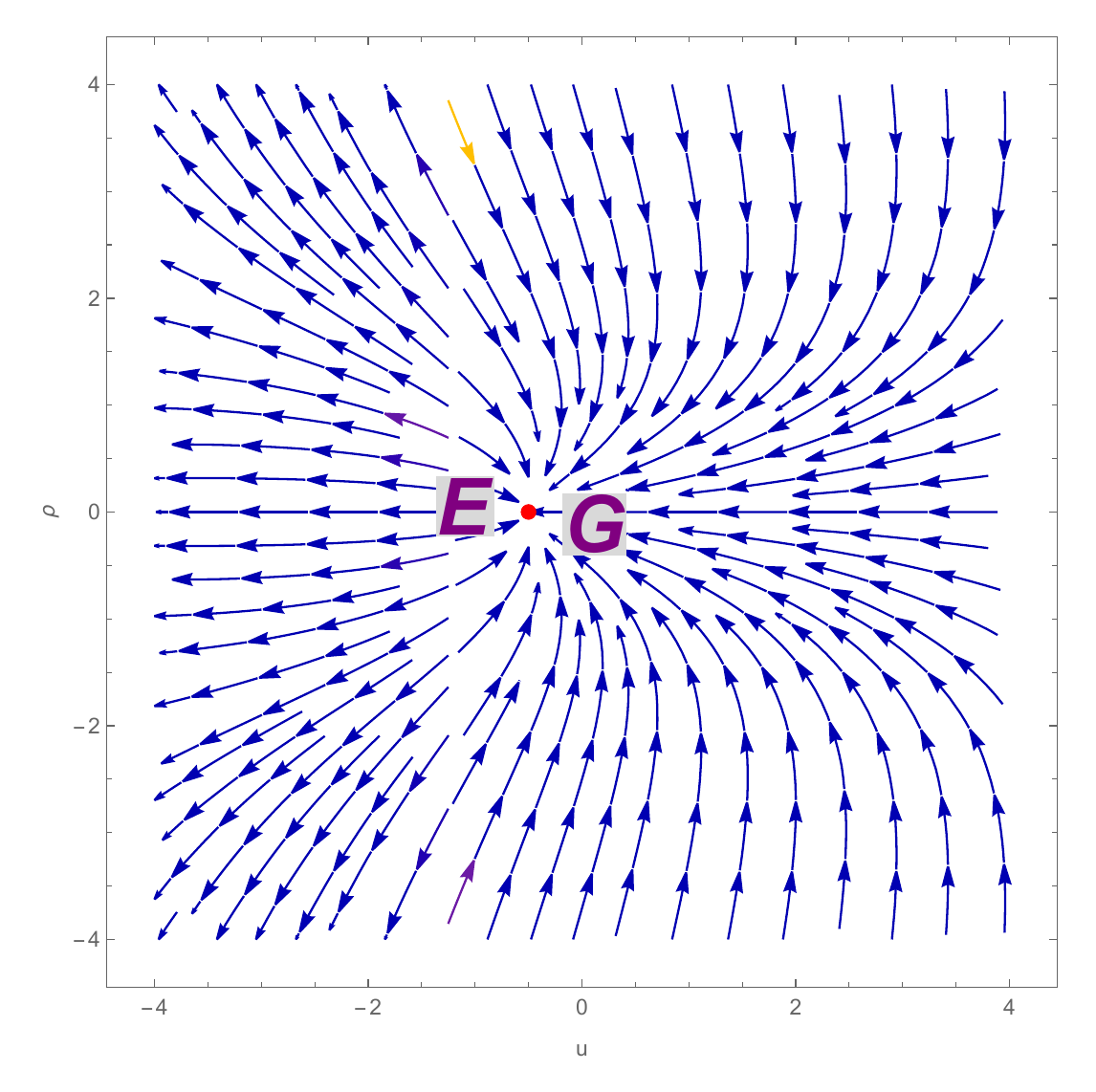}
    \includegraphics[width=58mm]{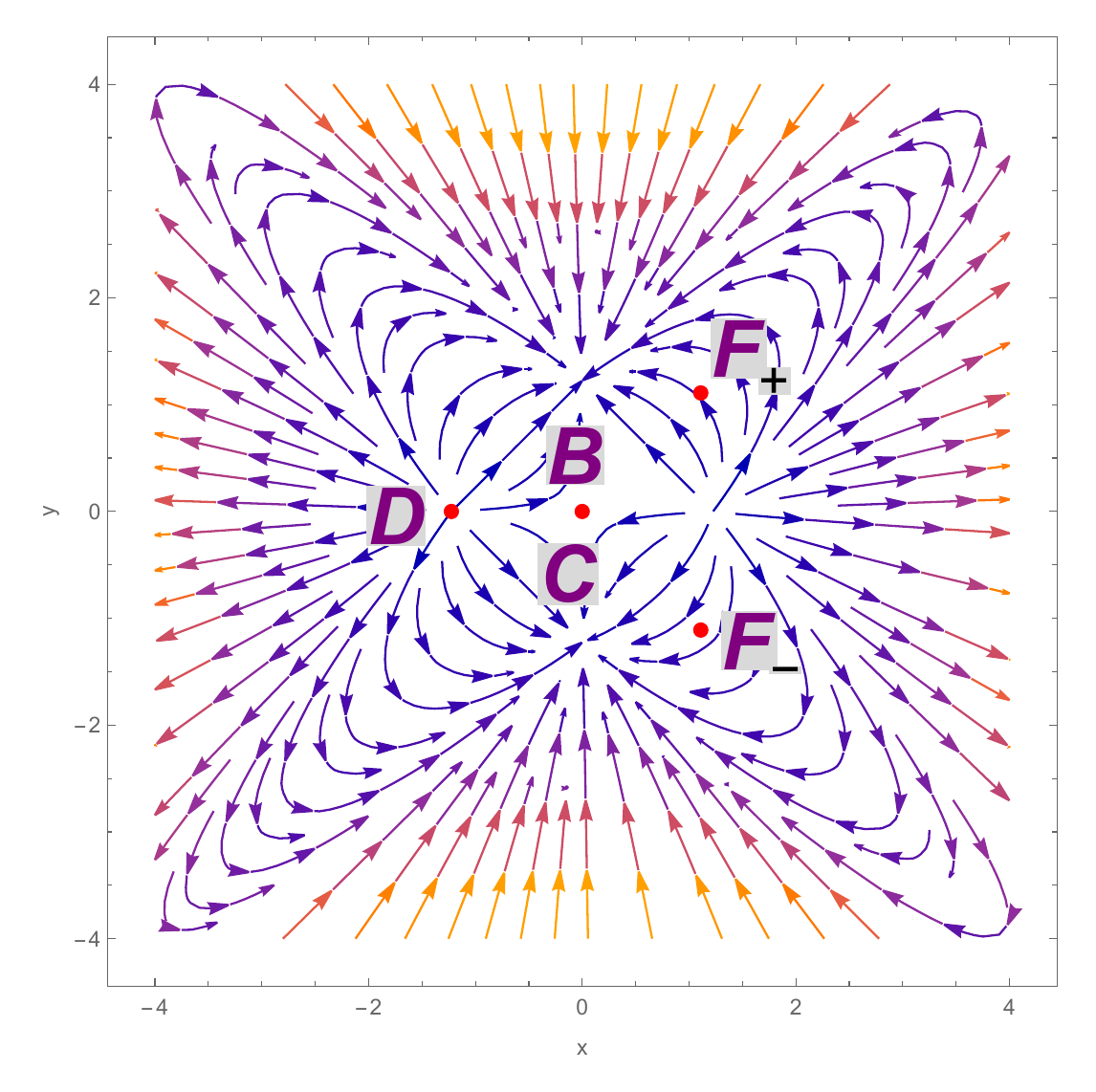}
    \caption{Phase portrait for the dynamical system of Model-I, (i) {\bf left panel}($x=0$, $y=0$, $z=1.5$, $\lambda=0.001$); (ii) {\bf middle panel} ($x=0$, $y=0$, $\lambda=0.001$) (iii) {\bf right panel} ($\rho=0$, $z=1.5$, $\lambda=0.001$.)} \label{Fig1}
\end{figure}

In the study of dynamical system, the phase portrait is an important tool, that consists of plot of typical trajectories in the state space. The stability of the models can be indicated through the phase portrait. Fig.~\ref{Fig1} shows the phase space portrait diagram for the dynamical system Eqs.~\eqref{36}--\eqref{41}. The {\bf left panel} shows that the trajectories of critical points $A$ and $\mathcal{H}$ move towards from the fixed point, so we conclude that the point $A$ and $\mathcal{H}$ are stable nodes. Similarly, phase portrait in {\bf middle panel} indicates that the trajectories of the critical points $E$ and $G$ move towards the fixed point, showing the stable behaviours. The trajectories for the critical points $B$, $C$, $D$ and $F_{+}$,$F_{-}$ move away from the fixed points as in the { \bf right panel}. Hence, these points are unstable (saddle). Further, we have described in details the corresponding cosmology for each critical points as below:

\begin{itemize}
    \item{\bf Critical Point $A$ :} At this point, $\Omega_{de}=1$, $\Omega_{m}=0$ and $\Omega_{r}=0$, i.e the Universe shows dark energy dominated phase. The corresponding EoS parameter $\omega_{tot}=-1$ and deceleration parameter $q=-1$ confirms the accelerated dark energy dominated Universe. The eigenvalues of this critical point is negative real part and zero. Coley and  Aulbach \cite{Coley:1999uh,aulbach1984continuous} have investigated that the dimension of the set of eigenvalues for non-hyperbolic critical points is one equal to the number of vanishing eigenvalues. As a result, the set of eigenvalues is normally hyperbolic, and the critical point associated with it is stable but cannot be a global attractor. In our case, the dimension of set of eigenvalue is one and only one eigenvalue vanishes. That means the dimension of a set of eigenvalues is equal to the number of vanishing eigenvalues. This critical point is consistent with recent observations and can explain current acceleration of the Universe. The behaviour of this critical point is a stable node.
    \begin{align}
        \{-3,-3,-2,0\}\,. \nonumber
    \end{align}
    
    \item{\bf Critical Point $B$:} 
    This point exists for $\gamma \neq 1$ and the corresponding deceleration parameter $q=\frac{1}{2}$ and EoS parameter $\omega_{tot}=0$. This behaviour of the critical point leads to the decelerating phase of the Universe. Also, density parameters $\Omega_{de}=\gamma$, $\Omega_r=0 $ and $\Omega_{m}=1-\gamma$. If we consider $\gamma=0$, the Universe shows the matter-dominated phase. The eigenvalues of the Jacobian matrix for this critical point are given below. The signature of the eigenvalues is both positive and negative that means it shows the unstable saddle behaviour. 
    \begin{align}
        \left\{-\frac{3}{2},\frac{3}{2},-\frac{1}{2},0\right\}\,.\nonumber
    \end{align}
    
    \item {\bf Critical Point $C$:} At this point, the deceleration parameter and EoS parameter are obtained to be $q=1$ and $\omega=\frac{1}{3}$, which demonstrates the decelerating phase of the Universe. The density parameters are: $\Omega_{de}=\sigma $, $\Omega_{r}=1-\sigma $ and $\Omega_{m}=0$. For the value of $\sigma=0$, the Universe exhibits radiation-dominated phase i.e. $\Omega_{r}=1 $. The eigenvalues of Jacobian matrix for this critical point are given below and since it contains both negative and positive eigenvalues, this critical point is an unstable saddle.
    \begin{align}
        \{-1,1,2,0\}\,. \nonumber
    \end{align}
    
    \item{\bf Critical Point $D$:} The value of density parameters for this point are, $\Omega_{m}=0 $, $\Omega_{r}=0 $, and $\Omega_{de}=1 $. The EoS and deceleration parameter are respectively shows the value $q=2$ and $\omega_{tot}=1$ and so the point behaves as stiff matter and showing the decelerating behaviour. The eigenvalues are obtained to be positive real part and zero. Due to presence of positive eigenvalue, this critical point is showing unstable behaviour.
    \begin{align}
        \left\{0,1,3,\frac{6-\sqrt{6}\delta \lambda}{2}\right\}\,.\nonumber
    \end{align}

    \item{\bf Critical Point $E$:} The density parameters are $\Omega_{m}=0 $,  $\Omega_{r}=0 $, and $\Omega_{de}=1 $, which indicates the the dark energy sector of the Universe. The deceleration parameter value $q=-1$ and the EoS parameter value $\omega_{tot}=-1 $ shows the accelerating behaviour of the Universe at this point. The negative and zero eigenvalues demonstrates the stable behaviour. At this point, the Universe shows the stability behaviour at the the accelerating dark energy phase.
    \begin{align}
        \{-3,-3,-2,0\}\,.\nonumber
    \end{align}
    
    \item {\bf Critical Point $F_{+}$:} This critical point exists for $\mu \neq 1$ and $\lambda \neq 0$. The decelerating behaviour has been observed since the value of deceleration parameter $q=\frac{1}{2}$ and the EoS parameter vanishes. The density parameters exhibit the value, $\Omega_{m}=1-\frac{3}{\lambda^{2}}-\mu $, $\Omega_{r}=0 $, and $\Omega_{de}=\frac{3}{ \lambda^{2}}+\mu $. For, $\lambda=1$ and $\mu=-3$, the critical point shows the matter-dominated era, else described a non-standard cold dark matter-dominated era with $\omega_{tot}=0$. This critical point is stable if it satisfies the stability condition of  Table~\ref{TABLE-II} otherwise, unstable saddle behaviour due to the presence of both positive and negative eigenvalues.
    \begin{align}
        \left\{0,-\frac{1}{2},\frac{3}{4} \left(-\frac{\sqrt{\lambda ^2 (\mu -1) \left(-7 \lambda ^2 (\mu -1)-24\right)}}{\lambda ^2 (\mu -1)}-1\right),\frac{3}{4} \left(\frac{\sqrt{\lambda ^2 (\mu -1) \left(-7 \lambda ^2 (\mu -1)-24\right)}}{\lambda ^2 (\mu -1)}-1\right)\right\}\,.\nonumber
    \end{align}
\item { \textbf{Critical Point $F_{-}$}: Similar to the critical point $F_{+}$ this critical point  exists for $\nu \neq 1$ and $\lambda \neq 0$. The value of deceleration parameter and the Eos parameter $\omega_{tot}$ are mentioned in Table~\ref{TABLE-II}. The density parameters values are, $\Omega_{m}=1-\frac{3}{ \lambda^{2}}-\nu $, $\Omega_{r}=0 $, and $\Omega_{de}=\frac{3}{\lambda^{2}}+\nu $. For, $\lambda=1$ and $\nu=-3$, the critical point indicates the matter-dominated period, else described a non-standard cold dark matter-dominated era with $\omega_{tot}=0$. From the stability criteria, it is clear that this critical point represents stable behaviour if it satisfies stability condition which is mentioned in Table~\ref{TABLE-II}. Otherwise, it exhibits unstable saddle behavior because both positive and negative eigenvalues are present.
    \begin{align}
        \left\{0,-\frac{1}{2},\frac{3}{4} \left(-\frac{\sqrt{\lambda ^2 (\nu -1) \left(-7 \lambda ^2 (\nu -1)-24\right)}}{\lambda ^2 (\nu -1)}-1\right),\frac{3}{4} \left(\frac{\sqrt{\lambda ^2 (\nu -1) \left(-7 \lambda ^2 (\nu -1)-24\right)}}{\lambda ^2 (\nu -1)}-1\right)\right\}\,.\nonumber
    \end{align}}\\
The definition of dimensionless variable $y$ as described in Eq.\eqref{27} allow us to study the different phases of the Universe evolution. The critical points with the condition on $y$ as if $y>0$ it correspond to the positive Hubble parameter and can explain the expanding universe. While the critical points with $y<0$ correspond to the $H<0$ describe the contracting phase of the universe \cite{dutta2018}. We denote the subscripts $+$ or $-$ corresponding to the
critical point $F$ with $y>0$ or $y<0$.
    \item{ \bf Critical Point $G$:} Here, we obtained $\Omega_{m}=0 $, $\Omega_{r}=0 $ and $\Omega_{de}=1 $, which shows the dark energy era of the evolution. The deceleration parameter value $q=-1$ confirms the accelerating behaviour whereas the EoS parameter value $\omega_{tot}=-1$ shows the $\Lambda$CDM like behaviour. The stability of the critical point has been confirmed from the eigenvalues.
    \begin{align}
        \{0,-3,-3,-2\}\,.\nonumber
    \end{align}
    
    \item{\bf Critical Point $\mathcal{H}$:} It describes the dark energy dominated phase as, $\Omega_{m}=0 $, $\Omega_{r}=0 $, and $\Omega_{de}=1 $. The accelerating behaviour and the EoS parameter depend on the relation of $\mathbf{i}$ and $\mathbf{j}$ as described in Table~\ref{TABLE-II}. For, $\mathbf{j}>2\mathbf{i}^{2}$, the deceleration parameter and EoS parameter exhibit the accelerating phase of the Universe. The eigenvalues, as given below indicate that there is a region in the parameter space where this point are stable nodes and attractor. Since this is a de-Sitter solution, the values of the parameter listed in Table~\ref{TABLE-II} will experience an accelerated expansion. The stability behaviour can be observed for $\mathbf{j}>2\mathbf{i}^{2}$.
    \begin{align}
        \left\{\frac{3 \mathbf{i}^2}{2 \mathbf{i}^2-\mathbf{j}},-\frac{\mathbf{i}^2-2 \mathbf{j}}{2 \mathbf{i}^2-\mathbf{j}},-\frac{3 \left(\mathbf{i}^2-\mathbf{j}\right)}{2 \mathbf{i}^2-\mathbf{j}},-\frac{3 \mathbf{j}^2}{\left(2 \mathbf{i}^2-\mathbf{j}\right)^2}\right\}\,. \nonumber
    \end{align}
\end{itemize}

\begin{figure}[H]
    \centering
    \includegraphics[width=74mm]{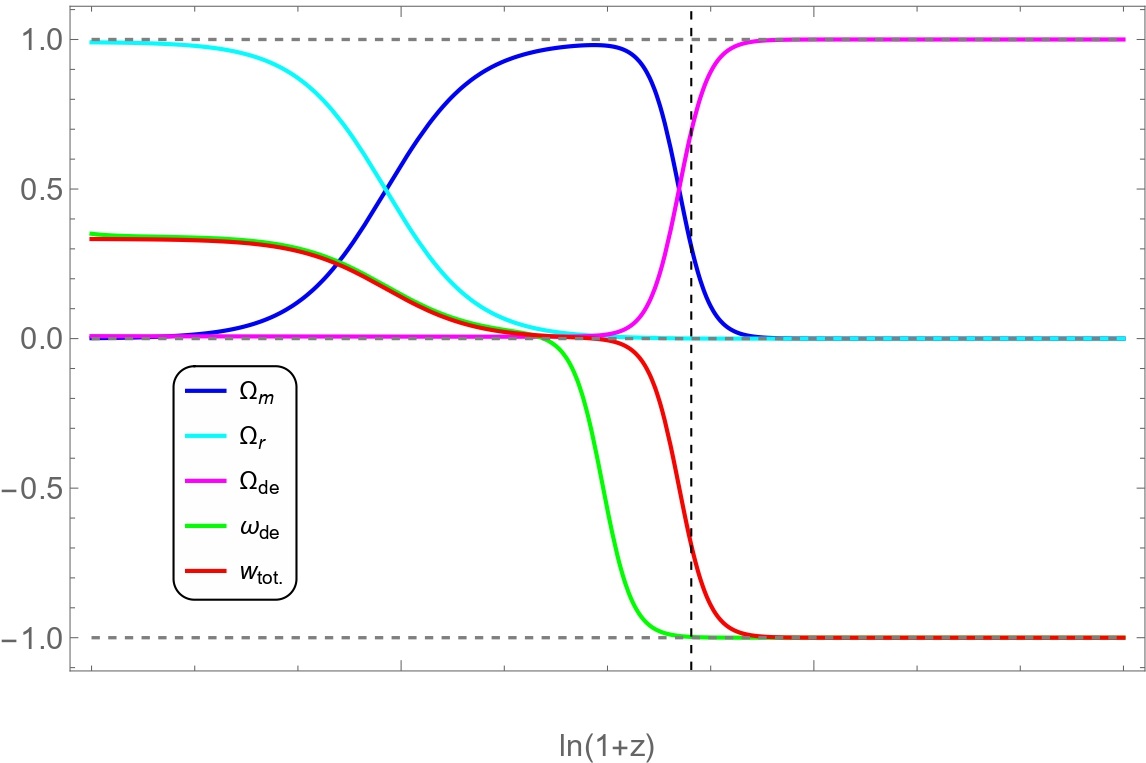}
    \caption{The evolution of the density parameters  as well as of the equation-of-state parameter, as functions of the redshift, for the case, $\lambda=0.001$ with the initial condition of dynamical system variables: $x=10^{-4}$, $y=10^{-6}$, $u=0.7 \times 10^{-2}$, $\rho=0.933254$, $z=10^{-4}$, which are representative for their definitions in Eqs.~(\ref{27},\ref{28}). The vertical
dashed line denotes the present cosmological time ($z=0$). } \label{Fig2}
\end{figure}

 The critical points $A$, $E$, $G$, and $\mathcal{H}$ are the last four attractors we found when dark energy was in charge, and the universe is accelerating. In addition, we have found that the critical points  $B$, $F_{+}$, and $F_{-}$ shows a matter-dominated phase, and point $C$ represents a radiation-dominated phase of the Universe and observed that the radiation and matter dominated critical points show unstable behavior. In Fig.~\ref{Fig2} we plot the behavior of the energy densities of dark energy, dark matter and radiation, as well as the total equation of state ($\omega_{tot.}$) and the equation of state of dark energy ( $\omega_{de}$) as functions of the redshift. Conveniently, we employ the redshift $z=\frac{a_{0}}{a} -1$ (with $a_{0} = 1$ as the current scale factor) as an independent variable. As is standard, $z = 0$ represents the present time of the Universe. The vertical dashed line in Fig.~\ref{Fig2} denotes the present cosmological time \cite{Bahamonde:2017ize}. In  Fig.~\ref{Fig2} we can observe that the cosmos is initially dominated by radiation, then transitions to dark matter dominance, and eventually ends up being dominated by dark energy. As mentioned above, the universe provides a scaling-accelerating solution, where the dark matter and dark energy density parameters remain around $0.3$ and $0.7$ respectively. Also, It is observed that the $\omega_{tot.} \approx -0.75$ and $\omega_{de} \approx -1$ at the current time $z = 0$, which is consistent with the
observational constraint from Planck data \cite{Planck:2018vyg}. In Fig.~\ref{Fig2} we can observe that, the Universe first dominated by the radiation era (Cyan curve), followed by a brief phase of matter dominance (Blue curve) and finally the cosmological constant (Pink curve). This behaviour of the density parameter indicates that the present Universe is dominated by dark energy. The EoS parameter (Red curve) begins with radiation at $\frac{1}{3}$, falls to $0$ during the matter-dominated period and finally rises to $-1$ leads to the $\Lambda$CDM model, which is a candidate for dark energy models.

\subsection{Model II}\label{sec:model_2}

In this case, we consider the form of $G(T)$ as, $G(T)= T+\alpha T^{2} $, where $\alpha$ is a constant \cite{Fortes:2021ibz}, which is a small generalization beyond TEGR.  For $\alpha=0$ the model reduces to the TEGR model. The Klein-Gordon equation in this case is same as in Eq.(\ref{22}) and for this $G(T)$, Eqs.~\eqref{17}-\eqref{18} become
\begin{eqnarray}
    \rho_{de}&=&\frac{\dot{\phi^{2}}}{2}+V(\phi)-T(1+3T \alpha)\,, \label{42}\\
    p_{de}&=&\frac{\dot{\phi^{2}}}{2}-V(\phi)+T(1+3 T \alpha)+4 \dot{H}(1+6 T \alpha)\,. \label{43}
\end{eqnarray}

To create an independent dynamical system, dimensionless variables can be specified through the following:
\begin{align}
    x=\frac{\kappa\dot{\phi}}{\sqrt{6}H}\,, \hspace{1cm} y=\frac{\kappa\sqrt{V}}{\sqrt{3}H}\,, \hspace{1cm}
    z=-2 \kappa^{2}\,,  \hspace{1cm} 
    u=-36 H^{2} \alpha \kappa^{2}\,, \label{44} \\ 
    \rho=\frac{\kappa\sqrt{\rho_{r}}}{\sqrt{3}H}\,,  \hspace{1cm} 
    \lambda= -\frac{V_{,\phi}(\phi)}{\kappa V(\phi)}\,,  \hspace{1cm} 
    \Theta= \frac{V(\phi) V_{,\phi \phi}}{V_{,\phi}(\phi)^{2}}\,. \label{45}
\end{align}

The dimensionless variables defined in Eq.~\eqref{44}--\eqref{45} also satisfy Eq.~\eqref{25}. The EoS parameter and deceleration parameter can be expressed in the form of dimensionless variable as,
\begin{align}
    q&= -1- \frac{3 x^{2}-3 y^{2}-3z-3u+3+\rho^{2}}{-2+2z+4u}\,, \label{46} \\
    \omega_{tot}&= -1- \frac{2(3 x^{2}-3 y^{2}-3z-3u+3+\rho^{2})}{3(-2+2z+4u)}\,, \label{47} \\
    \omega_{de} &= -\frac{3 \left(u+x^2-y^2\right)+\rho ^2 (2 u+z)}{3 (2 u+z-1) \left(u+x^2+y^2+z\right)}\,. \label{48}
\end{align}
Subsequently, the corresponding dynamical system can be obtained as,
\begin{align}
    \frac{dx}{dN}&=-\frac{x \left(\rho ^2-3 \left(u-x^2+y^2+z-1\right)\right)}{2 (2 u+z-1)}-3 x+\sqrt{\frac{3}{2}} \lambda y^2\,, \label{49} \\ 
    \frac{dy}{dN}&=-\frac{1}{2} y \left(\frac{\rho ^2-3 \left(u-x^2+y^2+z-1\right)}{2 u+z-1}+\sqrt{6} \lambda x\right)\,, \label{50} \\ 
    \frac{du}{dN}&=\frac{u \left(\rho ^2-3 \left(u-x^2+y^2+z-1\right)\right)}{2 (2 u+z-1)}\,, \label{51} \\
    \frac{d\rho}{dN}&=-\frac{\rho \left(\rho ^2+5 u+3 x^2-3 y^2+z-1\right)}{2 (2 u+z-1)}\,, \label{52} \\ 
    \frac{dz}{dN}&=0\,, \label{53} \\
    \frac{d\lambda}{dN}&= -\sqrt{6}(\Theta-1)x \lambda^{2}\,. \label{54}
\end{align}

Using the same approach as in Model I, the critical points of the autonomous dynamical system Eqs.(\ref{49}--\ref{54}) are listed in Table~\ref{TABLE-IV}. 

\begin{figure}[H]
    \centering
    \includegraphics[width=58mm]{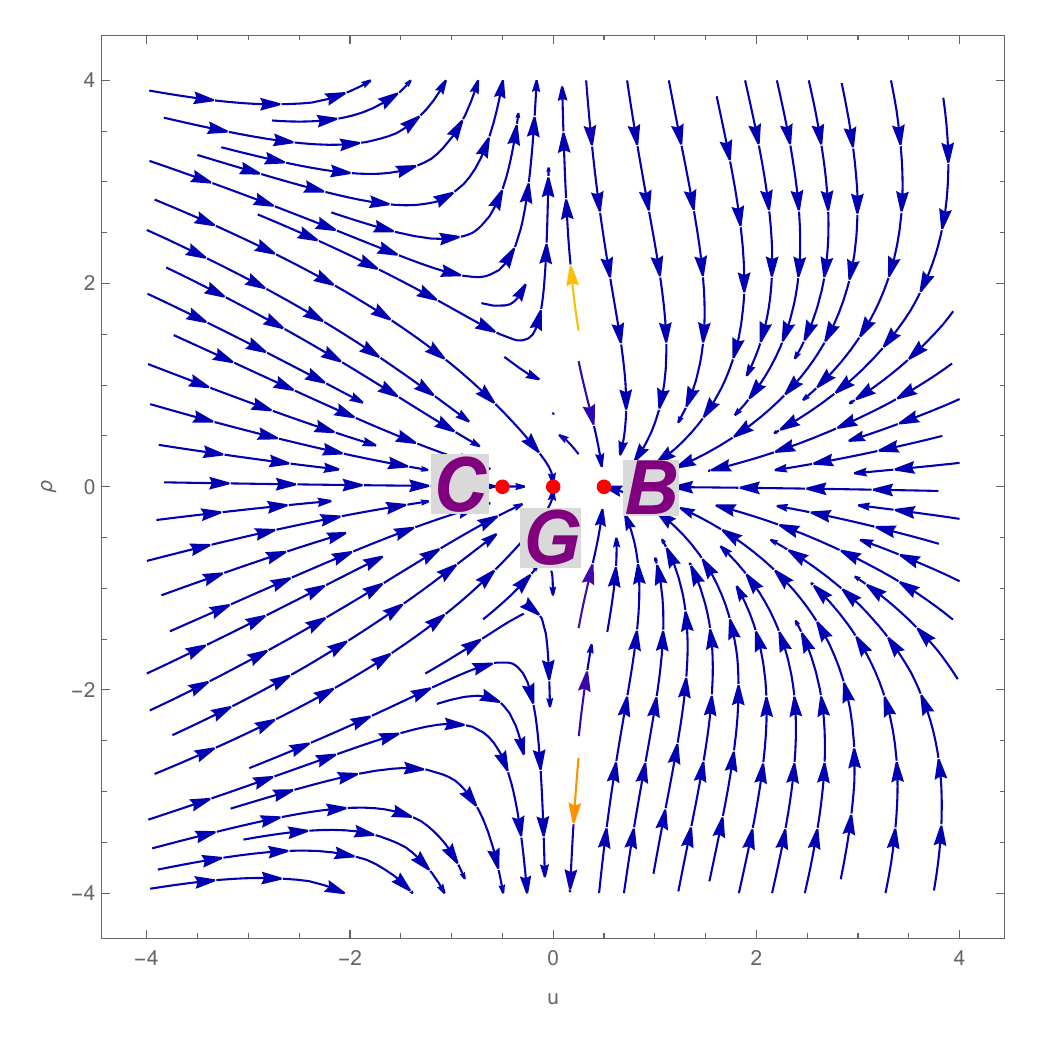}
    \includegraphics[width=58mm]{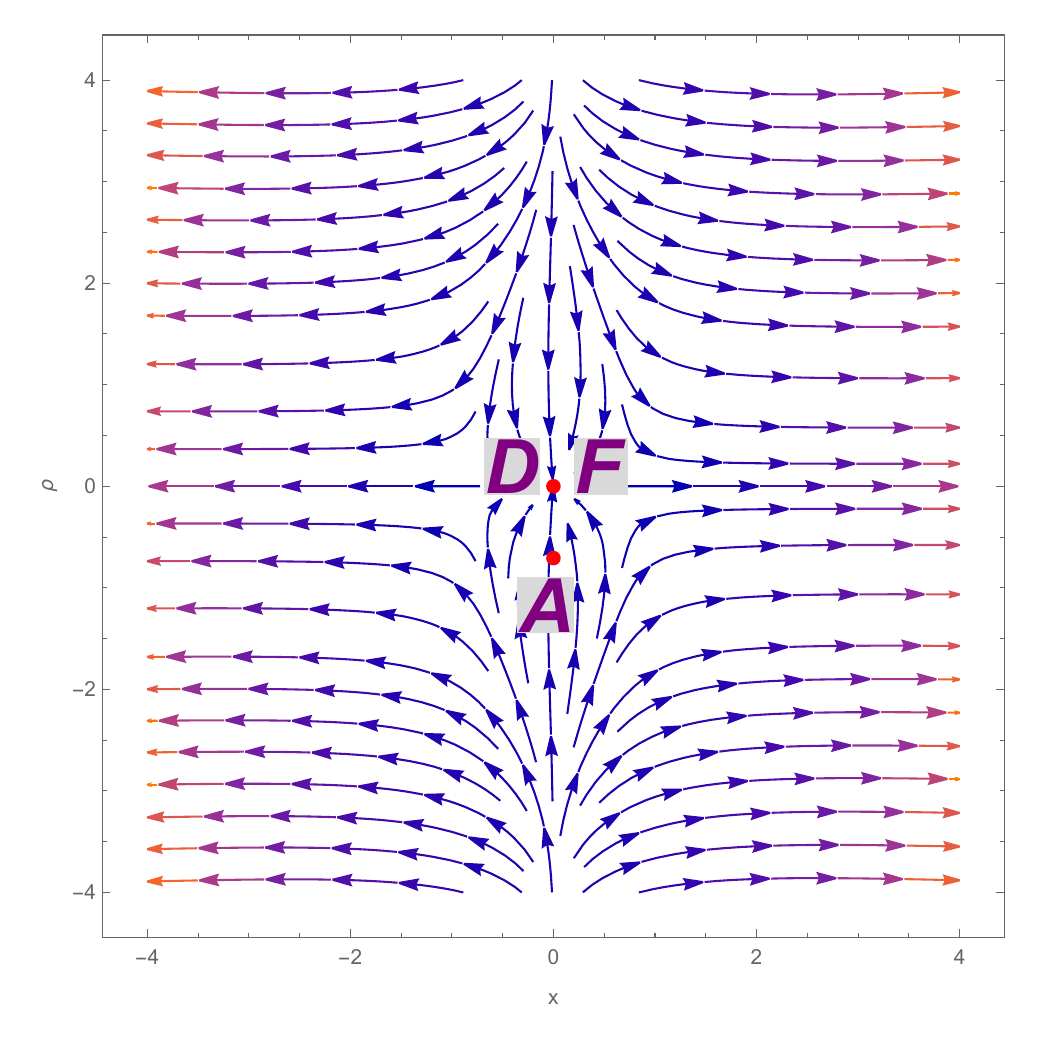}
    \includegraphics[width=58mm]{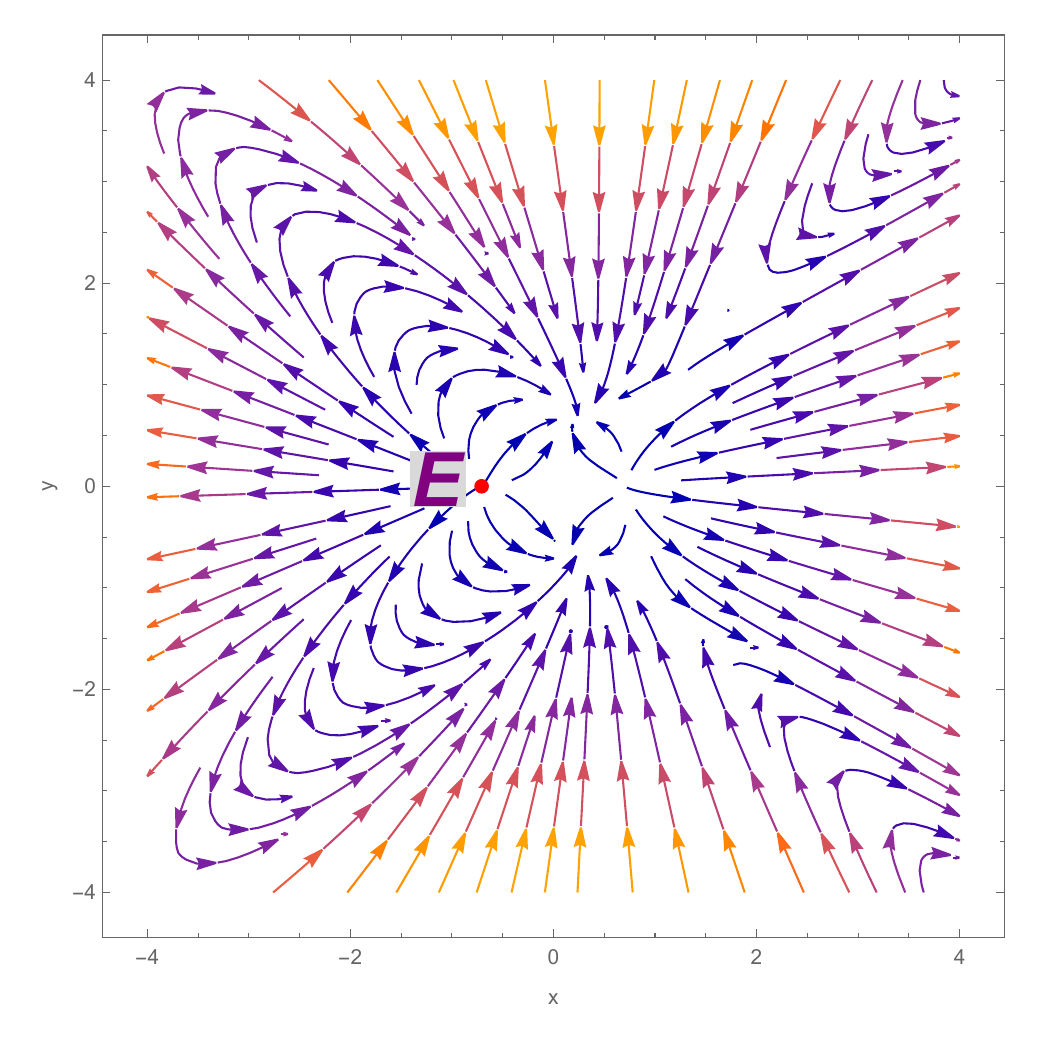}
    \caption{Phase portrait for the dynamical system of Model-II (i) {\bf left panel} ($x=0$, $y=0$, $z=0.5$, $\lambda=0.001$) ; (ii) {\bf middle panel} ($y=0$, $u=0$, $z=0.5$, $\lambda=0.001$) (ii) {\bf right panel} ($u=0$, $\rho=0$, $z=0.5$, $\lambda=0.001$).} \label{Fig3}
\end{figure}

\begin{table}[H]
    \caption{Critical Points for Dynamical System. } 
    \centering 
    \begin{tabular}{|c|c|c|c|c|c|c|} 
    \hline\hline 
    Critical Points & $x_{c}$ & $y_{c}$ & $u_{c}$ & $\rho_{c}$ & $z_{c}$ & Exists for \\ [0.5ex] 
    \hline\hline 
    $A$ & 0 & 0 & 0 & $\gamma_{1}$ & $\beta_{2}$ & $\begin{tabular}{@{}c@{}} $\gamma_{1}=-\sqrt{1-\beta_{2}}$,\\ $ \beta_{2}<1$ \end{tabular}$ \\
    \hline
    $B$ &0 & 0 & $\gamma_{2}$ & 0 &$\gamma $& $\begin{tabular}{@{}c@{}} $\gamma_{2}=1-\gamma$,\\ $-1+ \gamma \neq 0$ \end{tabular}$ \\
    \hline
    $C$ & 0 & $\xi$ & $\tau$ & 0 & $\sigma$ & $\begin{tabular}{@{}c@{}} $-1+2 \xi^{2}+\sigma \neq 0$,\\ $ \lambda=0$ \end{tabular}$ \\
    \hline
    $D$ & 0 & 0 & 0 & 0 & $\epsilon$ & $\epsilon \neq 1 $ \\
    \hline
    $E$ & $\gamma_{3}$ & 0 & 0 & 0 & $\alpha_{1}$ &  $\begin{tabular}{@{}c@{}}$\gamma_{3} =-\sqrt{1-\alpha_{1}},$\\ $ \alpha_{1}<1$ \end{tabular}$\\
    \hline
    $F$ & 0 & $\alpha_{2}$ & $\gamma_{4}$ & 0 & $\alpha_{3}$ & $\begin{tabular}{@{}c@{}}$\gamma_{4} =1-\alpha_{2}^{2}-\alpha_{3},$\\ $ -1+\alpha_{3} \neq 0,$ $\lambda \neq 0$ \end{tabular}$\\
    \hline
    $G$ & 0 & $\gamma_{5}$ & 0 & 0 & $\beta_{1}$ & $\begin{tabular}{@{}c@{}} $-1+\beta_{1} \neq 0,$\\ $\lambda=0$\end{tabular}$\\
    [1ex] 
    \hline 
    \end{tabular}
    \label{TABLE-IV}
\end{table}
For each critical point, the stability condition and to understand the corresponding cosmology, the deceleration and EoS parameter values are listed in Table~\ref{TABLE-V}. In Table~\ref{TABLE-VI}, the scale factor and the evolutionary phase of each critical point has been listed. Further to observe the stability behaviour of the critical points the phase portrait are given in Fig.~\ref{Fig3}. 
\begin{table}[H]
    \caption{Stability conditions, EoS Parameter and deceleration parameter } 
    \centering 
    \begin{tabular}{|c|c|c|c|c|} 
    \hline\hline 
    C. P. & Stability Conditions & $q$ & $\omega_{tot}$ & $\omega_{de}$ \\ [0.5ex] 
    \hline\hline 
    $A$ & \begin{tabular}{@{}c@{}}Stable for \\ $\frac{2}{5} < \beta_{2} < 1$ \end{tabular} & $1$ & $\frac{1}{3}$ & $\frac{1}{3}$ \\
    \hline
    $B$ & Stable &$-1$ &$ -1 $ & $-1$ \\
    \hline
    $C$ & Stable & $-1$ & $-1$ & $-1$ \\
    \hline
    $D$ & \begin{tabular}{@{}c@{}}Stable for \\ $\frac{2}{3} < \epsilon < 1$ \end{tabular}& $\frac{1}{2}$ & $0$ & $0$ \\
    \hline
    $E$ & Unstable & $2$ & $1$ & $1$ \\
    \hline
    $F$ & \begin{tabular}{@{}c@{}}Stable for \\ $\alpha _3>2 \alpha _2^2$ \end{tabular} & $\frac{-\alpha _2^2-2 \alpha _3+2}{-4 \left(-\alpha _2^2-\alpha _3+1\right)-2 \alpha _3+2}$ & $\frac{3 \left(-\alpha _2^2-\alpha _3+1\right)}{-6 \left(-\alpha _2^2-\alpha _3+1\right)-3 \alpha _3+3}$ & $\frac{\alpha _2^2+\alpha _3-1}{\left(\alpha _2^2-1\right) \left(2 \alpha _2^2+\alpha _3-1\right)}$ \\
    \hline
    $G$ & Stable & $-1$ & $-1$ & $-1$ \\
    [1ex] 
    \hline 
    \end{tabular}
    \label{TABLE-V}
\end{table}

\begin{table}[H]
    \caption{Cosmological solutions of critical points } 
    \centering 
    \begin{tabular}{|c|c|c|c|} 
    \hline\hline 
    C. P. & Acceleration equation & Scale factor(Power law solution) & Universe phase \\ [0.5ex] 
    \hline\hline 
    $A$ & $\dot{H}=-2 H^{2}$ & $a(t)= t_{0} (2 t+c_{2})^\frac{1}{2}$ & radiation-dominated \\
    \hline
    $B$ & $\dot{H}=0$ & $a(t)=t_{0} e^{c_{1}t}$ & de-sitter phase \\
    \hline
    $C$ & $\dot{H}=0$ & $a(t)=t_{0} e^{c_{1}t}$ & de-sitter phase \\
    \hline
    $D$ &$\dot{H}=-\frac{3}{2}H^{2}$ & $a(t)= t_{0} (\frac{3}{2}t+c_{2})^\frac{2}{3}$ & matter-dominated \\
    \hline
    $E$ & $\dot{H}=-2 H^{2}$ & $a(t)= t_{0} (3 t+c_{2})^\frac{1}{3}$ & stiff-matter   \\
    \hline
    $F$ & $\dot{H}=0$ & $a(t)=t_{0} e^{c_{1}t}$ & de-sitter phase \\
    \hline
    $G$ &$\dot{H}=0$ & $a(t)=t_{0} e^{c_{1}t}$ & de-sitter phase \\
    \hline 
    \end{tabular}
    \label{TABLE-VI}
\end{table}

The phase portrait diagram Fig.~\ref{Fig3} displays the critical points. Plots of these phase space trajectories are shown for the dynamical system indicated in Eqs \eqref{49}-\eqref{54}. The {\bf left panel} plot shows that the phase space trajectories are moving towards from critical points $B$, $C$, and $G$ hence these points represent stability with stable node point behaviour. If the critical points $A$, $D$ and $F$ satisfy the stability condition given in Table~\ref{TABLE-IV}, then phase space trajectories are moving towards the critical points $A$, $D $, and $F$. Otherwise phase portrait are moving away from these critical points {\bf middle panel}, we can observe that the critical point $A$, $D$, and $F$ are showing unstable behaviour. The {\bf right panel} phase portrait shows that the critical point $E$ trajectories deviate from the fixed point, indicating unstable behaviour. Additionally, we have included detailed descriptions of the associated cosmology at each critical point, below:

\begin{itemize}
    \item{\bf Critical Point A :} The density parameters for this point are $\Omega_{m}=0$, $\Omega_{r}= 1-\beta_{2}$ and $\Omega_{de}=\beta_{2}$. The behaviour depends on the value of the parameter $\beta_{2}$. For $\beta_{2}=0$, the critical point satisfies the radiation dominated phase. The positive deceleration parameter shows the decelerating phase of the Universe and the EoS parameter yields the value, $\omega_{tot}=\frac{1}{3}$. The eigenvalues for this critical point are given below, which can be interpreted as if the parameter $\beta$ satisfies the stability condition mentioned in Table~\ref{TABLE-IV}, then this critical point is stable, otherwise unstable.
    \begin{align}
        \left\{-\frac{\beta_{2} (\beta_{2} +2)}{(3 \beta_{2} -2)^2},-\frac{2}{3 \beta_{2} -2},\frac{4 (\beta_{2} -1)}{3 \beta_{2} -2},-\frac{5 \beta_{2} -2}{3 \beta_{2} -2}\right\}\,. \nonumber
    \end{align}
    
    \item{\bf Critical Point B :} Both the deceleration parameter and EoS parameter are showing the accelerating $\Lambda$CDM like behaviour. The dark energy phase has been confirmed from the density parameters, which are $\Omega_{m} = 0$, $\Omega_{r} = 0$ and $\Omega_{de} = 1$. The eigenvalues are either negative or zero, hence it confirms the stability behaviour.
    \begin{align}
        \{-3,-3,-2,0\}\,.\nonumber
    \end{align}
    
    \item{\bf Critical Point C :} Similar behaviour has been obtained for this point as in the critical point $B$, i.e. the accelerating $\Lambda$CDM like behaviour. The nature of the eigenvalues confirms the stability.
    \begin{align}
        \{-3,-3,-2,0\}\,.\nonumber
    \end{align}
    
    \item{\bf Critical Point D :} This point exists for $\epsilon \neq 1$. For this condition the vanishing EoS parameter shows the matter dominated Universe  with the deceleration parameter $q=\frac{1}{2}$. Hence the density parameters $\Omega_{m}=1-\epsilon$ and $\Omega_{de}=\epsilon$. From the eigenvalues of the critical point, we can conclude that for $\frac{2}{3}<\epsilon<1$, it shows the stability, else the unstable behaviour.
    \begin{align}
        \left\{-\frac{3 \epsilon ^2}{(3 \epsilon -2)^2},\frac{3 (\epsilon -1)}{3 \epsilon -2},-\frac{3 (2 \epsilon -1)}{3 \epsilon -2},-\frac{3 \epsilon -1}{3 \epsilon -2}\right\}\,.\nonumber
    \end{align}
    
    \item{\bf Critical Point E :} At this point, $\Omega_{m}=0$, $\Omega_{r}=0$ and $\Omega_{de}=1$ with $\omega_{tot}=1$ and $q=2$. The behavior of this critical point is always unstable due to the presence of positive and negative eigenvalues. At the point when dark energy dominates the Universe, the EoS parameter reduced to a stiff fluid and there is no sign of acceleration. 
    \begin{align}
        \left\{-\frac{2}{3 \alpha _1-2},\frac{-2 \sqrt{6} \sqrt{1-\alpha _1} \lambda +3 \sqrt{6} \alpha _1 \sqrt{1-\alpha _1} \lambda +12 \alpha _1-12}{2 \left(3 \alpha _1-2\right)}, \frac{3 \left(2+2 \alpha _1^2-5 \alpha _1-\sqrt{7 \alpha _1^4-28 \alpha _1^3+37 \alpha _1^2-20 \alpha _1+4}\right)}{\left(3 \alpha _1-2\right){}^2}\right. \nonumber \\
        \left.\frac{3 \left(2+2 \alpha _1^2-5 \alpha _1+\sqrt{7 \alpha _1^4-28 \alpha _1^3+37 \alpha _1^2-20 \alpha _1+4}\right)}{\left(3 \alpha _1-2\right)^{2}}\right\}\,. \nonumber
    \end{align}
    
    \item{\bf Critical Point $F$:} The solution to this critical point is $\Omega_{r}=0$, $\Omega_{de}=1- \alpha_{2}^{2}$ and $\Omega_{m}=\alpha_{2}^{2}$ with the EoS and and deceleration parameter are as in Table~\ref{TABLE-V}. The EoS parameter satisfying this condition $\alpha _3<1-2 \alpha _2^2$. It is interesting to note that in this case, the final value of $\omega_{tot}$ ranges between -$\frac{1}{3}$ to $-1$. For this condition, the EoS parameter and deceleration parameters indicate accelerated phase of the Universe. For $\alpha_{2}=0$, the critical point indicates a period where the Universe is dominated by dark energy era ($\Omega_{de}=1$). Also, the behavior of the EoS and deceleration parameters for $\alpha_{2}= 0$ shows an accelerating phase of the Universe. The critical point is stable for $\alpha _3>2 \alpha _2^2$ and for this condition, all the eigenvalues are negative which confirms the stability behaviour.
    \begin{align}
        \left\{\frac{3 \alpha _2^2}{2 \alpha _2^2-\alpha _3},-\frac{\alpha _2^2-2 \alpha _3}{2 \alpha _2^2-\alpha _3},-\frac{3 \left(\alpha _2^2-\alpha _3\right)}{2 \alpha _2^2-\alpha _3},-\frac{3 \alpha _3^2}{\left(2 \alpha _2^2-\alpha _3\right){}^2}\right\}\,. \nonumber
    \end{align}
    
    \item{\bf Critical Point $G$:} As the values of the density parameters, deceleration parameter and EoS parameter become same as that of the critical point $B$ and $C$ and also the eigenvalues, therefore the behaviour of this critical point $G$ remains same as that of $B$ and $C$.
    \begin{align}
        \{-3,-3,-2,0\}\,.\nonumber
    \end{align}
\end{itemize}

\begin{figure}[H]
    \centering
    \includegraphics[width=74mm]{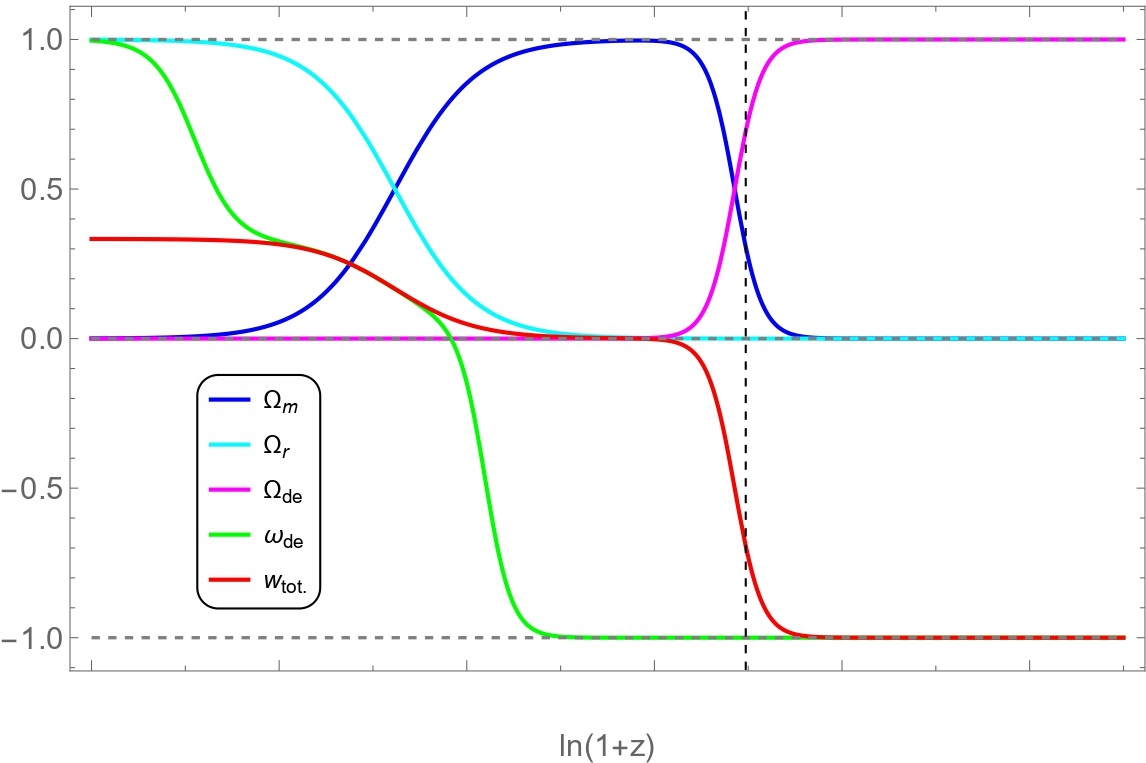}
    \caption{The evolution of the density parameters ($\Omega_{de}$), ($\Omega_{m}$) and, ($\Omega_{r}$)  as well as of the equation-of-state parameter, as functions of the redshift, $\lambda=0.001$ with the initial conditions of dynamical system variables: $x=10^{-6}$, $y=10^{-6}$, $u=10^{-15}$, $\rho=0.933234$, $z=10^{-8}$. The vertical
dashed line denotes the present cosmological time ($z=0$).} \label{Fig4}
\end{figure}

 The critical points $B$, $C$, $F$, and $G$ are representing the dark energy sector and showing late-time cosmic acceleration behaviour of the Universe. These critical points show the attractor phase (stable). The critical points $A$ and $D$ indicate the matter and radiation phase respectively and show unstable behaviour of the Universe. In Fig.~\ref{Fig4}, the evolution of the energy densities as well as EoS parameter as a function of redshift has been shown. The EoS parameter ($\omega_{tot}$) (Red curve) of the cosmos together with the relative energy densities of dark matter ($\Omega_{m}$), radiation ($\Omega_{r}$) and dark energy ($\Omega_{de}$) are shown. The evolution shows the radiation phase (Cyan curve), followed by a brief period of domination by the matter (Blue curve), and after that the domination of dark energy sector (Pink curve). We observe that the Universe transit from a matter dominated phase to an acceleration era at late times. The present value of the dark matter and dark energy density parameters remain respectively, around $0.3$ and $0.7$ at $z=0$. Also, we have found $\omega_{tot.}\approx -0.76$ and $\omega_{de}\approx -1$ at the present cosmic time. The EoS parameter approaches to $-1$ leading to the $\Lambda$CDM behaviour of the model. 

\section{Discussions and Conclusion} \label{SECIV}

The dynamical systems technique offers a crucial approach in the toolkit of probes of background cosmology. It offers an avenue to explore what critical points a model has associated with it, and what are the natures of each of these points. These points can then be correlated with the evolution of the Universe as evidenced from observational cosmology, which can be a compelling first test of any proposed model stemming from modified gravity. Moreover, the coupling of the critical points analysis together with their stability and eventual phase portraits can provide compounded evidence to support or reject particular models or parameter ranges within the selected models.

In this work, we explored the dynamical systems analysis of two particular models within the general class of scalar-tensor theories coupled with the torsion scalar, as prescribed in Eq.~\eqref{1}. The effective Friedmann and Klein-Gordon equations provided in Eqs.~\eqref{12}--\eqref{14} describe fully the background dynamics of the system, but are beyond analytic techniques and so we explore their dynamics using dynamical systems analysis. Models in this class of theories may offer some advantages such as the scalar field and torsion scalar freedoms being associated with different epochs of the evolution of the Universe, or with different mechanisms within the Universe.

The scalar field is ultimately described canonically with an exponential potential. On the other hand, building on the proposals in Ref.~\cite{2011JCAP...07..015Z}, we use logarithmic and power-law models to describe the form of the torsion scalar term beyond TEGR. These were first probed in an $f(T)$ gravity context in Ref.~\cite{2011JCAP...07..015Z} where they were found to have some advantageous properties which were correlated with the evolution of the Universe. Adding a scalar field may produce more realistic cosmology since scalar fields have been suggested to be responsible for a variety of different mechanisms in the Universe such as inflation and late-time accelerated expansion. In our analysis, we find that the logarithmic model developed in Sec.~\ref{sec:model_1} produces a rich cosmology as shown through the critical points in Table~\ref{TABLE-I} which are then further studies for the nature of their critical points in Table~\ref{TABLE-II}. To show these properties in fuller details, we also include phase portraits in Fig.~\ref{Fig1} where the behaviour at those points is more clearly represented. The behaviour of the scale factor at each critical point is shown in Table~\ref{TABLE-III}.  If we compare the analysis made in  Ref.~\cite{2011JCAP...07..015Z} for the logarithmic model, we can describe that, there are eight more critical points. The study made in  Ref.\cite{2011JCAP...07..015Z}, successfully explain the de-sitter solution through the dynamical system analysis of logarithmic model and conclude that this study will not explain radiation and matter dominated era of the universe evolution. The cosmology based on this study of logarithmic model along with the addition of scalar field successfully explain de-sitter solution in the matter and radiation dominated phases of the evolution of the Universe. In this study, we have added the scalar field to explain both the radiation and matter dominated era. We close the discussion with the figure that describes the evolutionary behaviour of various density parameters and EoS parameters.

In our second model, explore in Sec.~\ref{sec:model_2}, we take a square torsion scalar extension to the TEGR term. This would represent many other extensions as a leading order term in most circumstances such as background cosmology. Again, here we define suitable dynamical variables and provide the autonomous dynamical system in Eqs.~\eqref{49}--\eqref{54}. This leads to the critical points presented in Table~\ref{TABLE-IV} together with their properties as described in Table~\ref{TABLE-V}. Similarly, we describe the behaviour of each the scale factor at each critical point in Table~\ref{TABLE-VI}. Finally, the phase portraits of Fig.~\ref{Fig3} are shown where the nature of each critical point is shown more fully through the evolutionary contours. Finally, we close with a diagram showing the evolution of each density parameter in Fig.~\ref{Fig4}.

This work shows the promise of these two models which should be further explored more deeply in the cosmological context through observational constraint analysis, or through perturbation theory which may reveal more information about these models such as their links to the large scale structure of the Universe and the cosmic microwave background radiation power spectrum. It may also be interesting to study these models in different contexts such as in astrophysics either in the weak or string fields.

\section*{Acknowledgements} LKD acknowledges the financial support
provided by University Grants Commission (UGC) through
Junior Research Fellowship UGC Ref. No.: 191620180688 and SAK for the Senior Research Fellowship UGC Ref. No.: 191620205335 to carry out the research work. BM acknowledges the support of IUCAA, Pune (India) through the visiting
associateship program. JLS would like to acknowledge funding from Cosmology@
MALTA which is supported by the University of Malta. The authors would like to acknowledge support from the Malta Digital Innovation Authority through the IntelliVerse grant. The authors are thankful to the anonymous referees for their valuable comments and suggestions to improve the quality of the paper.

\section*{References}
\bibliographystyle{utphys}
\bibliography{references}

\end{document}